\documentclass[prb,aps]{revtex4}
\usepackage{epsfig,epsf,psfrag}
\usepackage{graphicx}
\usepackage{epic,eepic}
\usepackage{color,pstcol}
\begin{document}
\title{Positive noise cross-correlations in superconducting
hybrids: Role of interface transparencies}
\date{\today}

\author{R. M\'elin$^1$, C. Benjamin$^{2,3}$ and T. Martin$^{2,4}$}

\affiliation{$^1$ Institut NEEL,
  CNRS and Universit\'e Joseph Fourier, BP 166,
  F-38042 Grenoble Cedex 9, France}
\affiliation{$^2$ Centre de Physique Th\'eorique, Case 907 Luminy, 13288
  Marseille cedex 9, France}
\affiliation{$^3$ Quantum Information Group,
School of Physics and Astronomy,
University of Leeds,
Woodhouse Lane, Leeds,
LS29JT, UK}
\affiliation{$^4$ Universit\'e de la M\'edit\'erann\'ee, 13288 Marseille Cedex 9, France}


\begin{abstract}
Shot noise cross-correlations in normal
metal-superconductor-normal metal 
structures are discussed at
arbitrary interface transparencies using both the scattering approach of Blonder, 
Tinkham and Klapwik and a microscopic Green's function approach. 
Surprisingly,
negative 
crossed conductance in such set-ups [R. M\'elin and D. Feinberg,
Phys. Rev. B {\bf 70}, 174509 (2004)]
does not preclude the possibility of positive
noise cross-correlations for almost transparent contacts.
We conclude with a phenomenological discussion of interactions
in the one dimensional leads connected to the superconductor,
which induce sign changes in the noise
cross-correlations.
\end{abstract}
\maketitle

\section{Introduction}

Among the challenges in condensed matter systems at the nanoscale
is the manipulation of electronic entanglement in connection with
quantum information processing. Some theoretical proposals
\cite{Byers,Martin,Choi,Deutscher} involving superconductivity
have been implemented by three groups
\cite{Beckmann,Russo,Cadden}, aiming at the observation of non-local 
effects in transport, and in the long run at
the realization of a source of entangled pairs of electrons. The
devices realized up to now \cite{Beckmann,Russo,Cadden} consist of
ferromagnet-superconductor-ferromagnet (F$_a$SF$_b$) and normal
metal-superconductor-normal metal (N$_a$SN$_b$) three-terminal
hybrids, designed in such a way that Cooper pairs from the
superconductor have the opportunity to split in the two different
normal (N$_{a,b}$) or ferromagnetic (F$_{a,b}$) electrodes. Noise
measurements, on the other hand, have focused mostly on
two-terminal devices such as
a two dimensional electron
gas-superconductor junction\cite{brchoi-roche}
and a junction between a diffusive normal metal and
superconductor\cite{jehl-prober}.

The possibility of separating pairs of electrons into different
electrodes \cite{Lambert,Jedema}
has aroused a considerable theoretical interest
recently, and calls for a new look at fundamental issues related
to non local transport through a superconductor, namely tunneling
in three-terminal configurations in the presence of a condensate
\cite{Melin-wl,Falci,Samuelsson,Prada,Koltai,japs,Feinberg-des,ref-on-noise,%
Melin-Feinberg-PRB,Levy,Duhot-Melin,DM-EPJB,Morten,Giazotto,Golubov}.
In most of the above studies of three-terminal hybrids,
the leads are separately connected to
the source of Cooper pairs (the superconductor). Some early
proposals considered a normal metal ``fork'' with two leads each
connected to voltage probes, and with the third lead connected to
a superconductor. There, the motivation was to compute the current
noise cross-correlations between the two normal leads, and to see whether
the sign of this signal would be reversed compared to the negative
noise cross-correlations of a normal metal three-terminal device
(the fermionic analog of the Hanbury Brown and Twiss experiment).
Ref.[\onlinecite{martin_torres}] considered a ballistic fork with
an electron beam splitter, and found that positive noise
correlations could arise upon reducing the transparency of the
superconductor interface. Ref.[\onlinecite{Samuelsson}]
considered a chaotic cavity connected to the superconductor and to
the leads, and found positive noise correlations which are
enhanced by the backscattering of the contacts, and are robust
when the proximity effect in the dot is destroyed by an external
magnetic field. Positive
correlations may also occur in the absence of correlated
injection\cite{cottet-texier}.

\begin{figure}[t]
\centerline{\includegraphics[width=8cm]{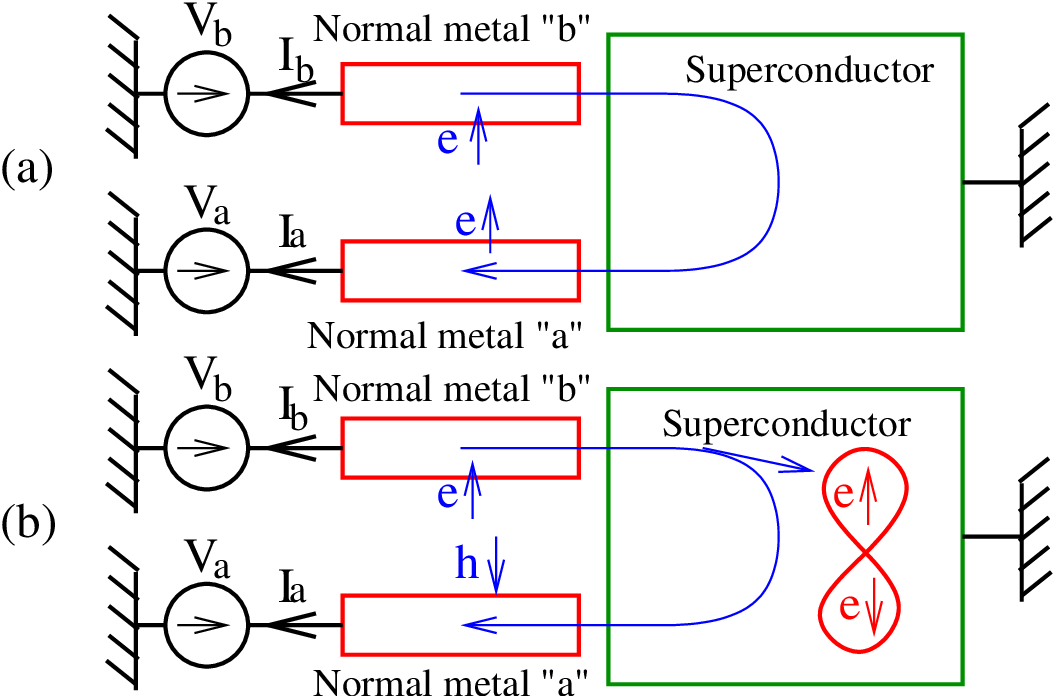}}
\caption{Schematic representation of elastic cotunneling (a) in a
N$_a$SN$_b$ structure, in which a spin-up electron (e$\uparrow$)
tunnels from electrode N$_b$ to electrode N$_a$ across the
superconductor. (b) shows crossed Andreev reflection in which a
spin-up electron (e$\uparrow$) coming from the normal electrode
N$_b$ is converted as a hole in the spin-down band (h$\downarrow$)
in electrode N$_a$ while a pair is transfered in the
superconductor. The balance between (a) and (b) can be controlled
by the relative spin orientation if the normal metals N$_{a,b}$
are replaced by ferromagnets F$_{a,b}$. The voltage $V_a$ is set
to zero in the available crossed conductance experiments
\cite{Beckmann,Russo,Cadden}. We evaluate in the article
cross-correlations between the currents $I_a$ and $I_b$ in the
presence of arbitrary voltages $V_a$ and $V_b$ in the normal or
ferromagnetic electrodes \cite{Bignon}. \label{fig:schema1} }
\end{figure}

It was shown already \cite{Martin,Choi} that noise
cross-correlations between the currents in different electrodes can lead
in the long term to Einstein, Podolsky, Rosen experiments with
electrons, being massive particles. In the short term,
current-current correlations among electrodes N$_a$ and N$_b$ in
N$_a$SN$_b$ three-terminal
structures may lead to detailed informations about the
microscopic processes mediating non local transport. 
A spin-up
electron from electrode N$_b$ may be transmitted as a spin-up
electron in electrode N$_a$ across the superconductor (a channel
called as ``elastic cotunneling''), or it may be transmitted as a
spin-down hole in electrode N$_a$ with a pair left in the
superconductor (a channel called as ``crossed Andreev
reflection'') (see Fig.~\ref{fig:schema1}). However, in
N$_a$SN$_b$ structures with tunnel contacts, the elastic
cotunneling current of spin-up electrons in electrode N$_a$ turns
out to be opposite to the crossed Andreev reflection current of
spin-down holes in electrode N$_a$: as shown by Falci {\it et al.}
\cite{Falci} the lowest order contribution
to the crossed conductance is vanishingly small in a
N$_a$ISIN$_b$ structure, where an insulating layer I is supposed
to be inserted in between the normal and superconducting
electrodes. A finite crossed conductance is however restored to
next orders in the tunnel amplitudes
\cite{Melin-Feinberg-PRB,Duhot-Melin,DM-EPJB}, if the normal
electrodes N$_{a,b}$ are replaced by ferromagnets \cite{Falci}
F$_{a,b}$, or if interactions in the superconductor 
are taken into account \cite{Levy}.
\begin{figure*}
\begin{center}
\includegraphics [width=.8 \linewidth]{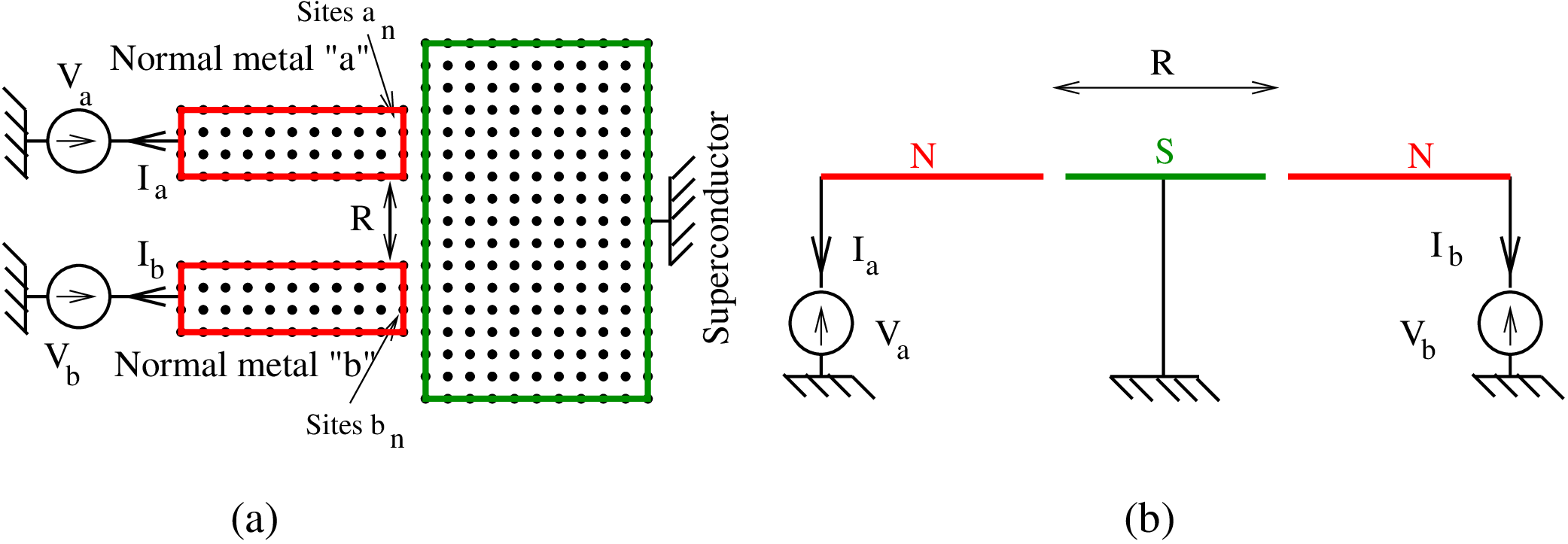}
\end{center}
\caption{Schematic representation of the tight-binding model of
N$_a$SN$_b$ structure with lateral contacts used in Green's function
calculations (a), and of the one dimensional (1D) N$_a$SN$_b$
structure used in the BTK approach (b).
\label{fig:schema2}
}
\end{figure*}

We are motivated by the fact that contacts with moderate to large 
interface transparency may be
used in future experiments in order to maximize the measured
current noise signal. A relevant question is then: do basic properties
such as the sign of current cross-correlations depend or not on
interface transparency, as compared to lowest order perturbation
theory in the tunnel amplitudes discussed by Bignon {\it et al.}
\cite{Bignon}? Interestingly, we obtain below a
positive answer to this question, using two complementary approaches: 
the scattering approach of Blonder Tinkham and Klapwijk as well as a 
microscopic tight binding scheme using the Keldysh method. 
The two approaches differ in the way interfaces are treated.
The scattering approach assumes that the N$_a$SN$_b$ system has a 
single mode connected throughout, while in the Green's function approach
the local tunneling term connects ``many'' modes in the leads to ``many''  
other modes in the superconductor. The study of these 
two limits is necessary to establish the robustness of the positive 
noise cross-correlation signal discussed in this paper. As we shall see, 
we find qualitative agreement between the two approaches.

Concerning the non local conductance calculation, the situation is summarized
rapidly. It was shown already
\cite{Melin-Feinberg-PRB,Duhot-Melin,DM-EPJB}
that the non local conductance of a N$_a$SN$_b$ structure is {\it negative}
for transparent interfaces. This was interpreted in terms of
the next higher order process in the tunnel transparencies \cite{Duhot-Melin}
corresponding to
the co-tunneling of pairs from electrode N$_b$ to electrode N$_a$.
Such processes is not expected to result in entanglement
in between the two electrodes N$_a$ and N$_b$ because no Cooper pair
from the superconductor is split between the normal electrodes
N$_a$ and N$_b$. 

However, noise crossed correlations constitutes a quantity
which is distinct from the non local conductance, and,
surprisingly, we show that the corresponding noise cross-correlations of
highly transparent N$_a$SN$_b$ structures 
are {\it positive} in spite of {\it negative} non local
conductance. This is confirmed by keeping track of the
processes contributing to the noise cross-correlations
in the Anantram and Datta \cite{ref-on-noise}
formula for the noise in multiterminal superconducting hybrids.

Finally, we mention that for experimental and practical reasons, the calculations 
presented in this paper are performed with the assumption that the distance which 
separates the two normal metal/superconductor interfaces is comparable or 
larger than the superconducting coherence length. Indeed, all experiments 
carried out so far for the non local conductance\cite{Beckmann,Russo,Cadden} 
have been performed within this range of parameters. 
For the BTK approach, this condition is necessary in order to avoid the
meaningless situation where all dimensions of the superconductor are
smaller than the coherence length. In the Green's function approach,
lateral contacts can be at a distance smaller than the coherence length
while all dimensions of the superconductor are much larger than the
coherence length. We verified carefully that our results are also
valid in this situation.

The article is organized as follows. Preliminaries 
for the Green's function and for the BTK approach are presented
in Sec.~\ref{sec:prelim}. The results of the two methods are discussed in
Sec.~\ref{sec:results}. Interactions in the normal metal leads are addressed in Sec.~\ref{sec:interactions}.

\section{Preliminaries}
\label{sec:prelim}

\subsection{Noise cross-correlations}
\label{sec:MGF}
The noise cross-correlation is given by the current fluctuations
\begin{equation}
\label{eq:cross}
{\cal S}_{a,b}(t,t')= \langle \delta I_a(t+t') \delta I_b(t)
\rangle ,
\end{equation}
where $I_a(t+t')$ and $I_b(t)$ are the currents through the normal
electrodes N$_{a,b}$ (or ferromagnetic electrodes F$_{a,b}$) (see
the notations ``a'' and ``b'' on Figs.~\ref{fig:schema1}
and~\ref{fig:schema2}) at times $t+t'$ and $t$ respectively.
The current operator at time $t$ is
defined as
\begin{equation}
I_a(t) = t_{a,\alpha} \sum_\sigma \sum_n c^+_{\alpha_n,\sigma}(t)
c_{a_n,\sigma}(t)+ h.c.
,
\end{equation}
where $h. c.$ is the Hermitian conjugate, $\langle ... \rangle$
denotes a quantum mechanical expectation value,
$c_{a_n,\sigma}(t)$ destroys a spin-$\sigma$ electron at time $t$
and site ``$a_n$'', and $c^+_{\alpha,\sigma}(t)$ creates a
spin-$\sigma$ electron at time $t$ and site ``$\alpha_n$'' (see
Fig.~\ref{fig:schema2} for the set of tight-binding sites labeled
by $a_n$, and for their counterpart $\alpha_n$ in the
superconducting electrode).

For DC bias applied to the electron reservoirs, time translational
invariance imposes that the real time cross-correlation ${\cal
S}_{a,b}(t_2-t_1)$ depends only on the difference of times of the
two current operators. This correlator is related in a standard
fashion to the Green's functions\cite{Caroli,Cuevas-noise}
$\hat{G}^+(t_2-t_1)$ and $\hat{G}^-(t_2-t_1)$ connecting the two
branches of the Keldysh contour:
\begin{eqnarray}
\label{eq:Sab}
{\cal S}_{a,b}&=& \left(\frac{e}{\hbar}\right)^2
\mbox{Tr}\left[
-\hat{G}^+_{b,\alpha}\otimes \hat{t}_{\alpha,a} \otimes
\hat{G}^-_{a,\beta} \otimes \hat{t}_{\beta,b} \right.\\
\nonumber
&&+ \hat{G}^+_{\beta,\alpha}\otimes
\hat{t}_{\alpha,a} \otimes \hat{G}^-_{a,b}
\otimes \hat{t}_{b,\beta}\\
\nonumber
&&+ \hat{t}_{\beta,b} \otimes \hat{G}^+_{b,a}\otimes
\hat{t}_{a,\alpha} \otimes \hat{G}^-_{\alpha,\beta} \\
&&- \left. \hat{G}^+_{\beta,a}\otimes
\hat{t}_{a,\alpha} \otimes \hat{G}^-_{\alpha,b}
\otimes \hat{t}_{b,\beta}
\right]
\nonumber
,
\end{eqnarray}
where the trace is a sum over the channels in real space (see
Fig.~\ref{fig:schema2}) and over electron and hole components of
the Nambu representation. The symbol $\otimes$ denotes 
convolution over time variables and the dependence on time variables
is implicit in Eq.~(\ref{eq:Sab}). The notations
$\alpha_n,\beta_n$ are used for the counterparts in the
superconductor of the tight-binding sites $a_n$ and $b_n$ at the
interfaces of electrodes N$_{a,b}$ (see Fig.~\ref{fig:schema2}).
The notation $\hat{g}_{i,j}$ corresponds to the Nambu Green's
function of electrodes isolated from each other, while the
notations $\hat{G}^+_{i,j}$ and $\hat{G}^-_{i,j}$ are used for
electrodes connected to each other by arbitrary interface
transparencies. For instance one has the following for the Keldysh
Green's function $G^+_{i,j}(t_1,t_2)$:
\begin{equation}
\hat{G}^+_{i,j}(t_1,t_2)=i \left(\begin{array}{cc}
\langle c_{j,\uparrow}^+(t_2) c_{i,\uparrow}(t_1) \rangle &
\langle c_{j,\downarrow}(t_2) c_{i,\uparrow}(t_1) \rangle \\
\langle c_{j,\uparrow}^+(t_2) c_{i,\downarrow}^+(t_1) \rangle &
\langle c_{j,\downarrow}(t_2) c_{i,\downarrow}^+(t_1) \rangle
\end{array} \right)
.
\end{equation}
In Eq.~(\ref{eq:Sab}), $\hat{t}_{i,j}$ denotes the
Nambu hopping amplitude from $i$ to $j$:
\begin{equation}
\hat{t}_{i,j}=\left(\begin{array}{cc}
|t_{i,j}| & 0 \\
0 & -|t_{i,j}| \end{array} \right)
.
\end{equation}

Based on a standard description \cite{Melin-Feinberg-PRB}, we
suppose that propagation in the normal electrodes N$_{a,b}$ is
local, as for a vanishingly small phase coherence length. 
Some processes that delocalize in the
direction parallel to the interfaces \cite{Melin-wl}
are not accounted for. Then the
multichannel contact on Fig.~\ref{fig:schema2} can be replaced by
a single channel contact, and this is why we use in
Eq.~(\ref{eq:Sab}) the notations $a$, $\alpha$, $\beta$, $b$
instead of $a_n$, $\alpha_n$, $\beta_n$ 
and $b_n$ (see Fig.~\ref{fig:schema2}a),
and average over the
Fermi phase factor $k_F R$ entering non local propagation in the
superconductor.

To discuss both highly transparent interfaces and an arbitrary
distance between the contacts, we solve the Dyson equation of the
form
\begin{eqnarray}
\hat{A} \hat{G}_{\alpha,\beta}&=&\hat{g}_{\alpha,\beta}
+\hat{X}_{\alpha,\beta} \hat{G}_{\beta,\beta} \\
\hat{B} \hat{G}_{\beta,\beta}&=&\hat{g}_{\beta,\beta}
+\hat{X}_{\beta,\alpha} \hat{G}_{\alpha,\beta}
,
\end{eqnarray}
with
\begin{eqnarray}
\hat{A}&=&\hat{I}-\hat{g}_{\alpha,\alpha}\hat{t}_{\alpha,a}
\hat{g}_{a,a} \hat{t}_{a,\alpha}\\
\hat{B}&=&\hat{I}-\hat{g}_{\beta,\beta}\hat{t}_{\beta,b}
\hat{g}_{b,b} \hat{t}_{b,\beta}\\
\hat{X}_{\alpha,\beta}&=&\hat{g}_{\alpha,\beta}\hat{t}_{\beta,b}
\hat{g}_{b,b}\hat{t}_{b,\beta}\\
\hat{X}_{\beta,\alpha}&=&\hat{g}_{\beta,\alpha}\hat{t}_{\alpha,a}
\hat{g}_{a,a}\hat{t}_{a,\alpha}
.
\end{eqnarray}
Then we have
\begin{eqnarray}
\hat{G}_{\alpha,\beta}&=& \hat{Y}^{-1}\hat{g}_{\alpha,\beta}
+\hat{Y}^{-1}\hat{X}_{\alpha,\beta} \hat{B}^{-1}\hat{g}_{\beta,\beta} \\
\hat{G}_{\beta,\beta}&=& \hat{B}^{-1}\hat{g}_{\beta,\beta}
+ \hat{B}^{-1}\hat{X}_{\beta,\alpha} \hat{G}_{\alpha,\beta}
,
\end{eqnarray}
where we used the notation
\begin{equation}
\hat{Y}=\hat{A}-\hat{X}_{\alpha,\beta}\hat{B}^{-1}\hat{X}_{\beta,\alpha}
.
\end{equation}
The Keldysh Green's functions $\hat{G}^{+,-}_{a,\alpha}(\omega)$
and $\hat{G}^{+,-}_{\alpha,a}(\omega)$
are next obtained from the Dyson-Keldysh equations\cite{Caroli,Cuevas}
\begin{eqnarray}
\nonumber
\hat{G}^{+,-}_{a,\alpha}(\omega)&=&
\left(\hat{I}+\hat{G}^R_{a,\alpha}(\omega)\hat{t}_{\alpha,a}
\right)\hat{g}^{+,-}_{a,a}(\omega)\hat{t}_{a,\alpha}
\hat{G}^A_{\alpha,\alpha}(\omega)\\
\label{eq:toto1}
&+& \hat{G}^R_{a,\beta}(\omega)\hat{t}_{\beta,b}\hat{g}^{+,-}_{b,b}(\omega)
\hat{t}_{b,\beta} \hat{G}^A_{\beta,\alpha}(\omega)\\
\nonumber
\hat{G}^{+,-}_{\alpha,a}(\omega)&=&
\hat{G}^R_{\alpha,\alpha}(\omega)\hat{t}_{\alpha,a}
\hat{g}^{+,-}_{a,a}(\omega)
\left(\hat{I}+\hat{t}_{a,\alpha}\hat{G}^A_{\alpha,\alpha}(\omega)\right)\\
\label{eq:toto2}
&+& \hat{G}^R_{\alpha,\beta}(\omega)\hat{t}_{\beta,b}
\hat{g}^{+,-}_{b,b}(\omega)
\hat{t}_{b,\beta} \hat{G}^A_{\beta,\alpha}(\omega)
,
\end{eqnarray}
where we Fourier transformed from time $t$ to frequency $\omega$. Note that
$\omega$ is conjugate to the time difference in Eq.~(\ref{eq:cross}), and
thus we consider zero frequency noise.
The notations $\hat{G}_{i,j}^{A,R}(\omega)$ are used for the advanced and
retarded fully dressed Green's functions connecting $i$ and $j$
at frequency $\omega$.

The method discussed here allows to treat an arbitrary distance
between the normal electrodes and contains
all information about quantum interference effects
\cite{Melin-PRB}, unlike quasiclassics \cite{Zaikin}.
We did not find qualitative changes
of the crossed conductance and of the sign of noise
cross-correlations when crossing over from $R\agt \xi$ 
to $R\alt \xi$
within the Green's function approach
and therefore we present the results only for $R\agt\xi$,
in which case we recovered for the crossed conductance
the behavior of
Refs.~\onlinecite{Melin-Feinberg-PRB,Duhot-Melin,DM-EPJB}.

\subsection{Blonder Tinkham Klapwijk (BTK) approach to noise cross-correlations}

The Blonder, Tinkham, Klapwijk (BTK\cite{BTK}) approach was previously
applied to non local transport~\cite{DM-EPJB}, i.e. computing
the current voltage characteristics in the different leads (see
also Ref.~\onlinecite{japs}). Here it is extended to address noise
cross-correlations. Considering a one dimensional (1D)
N$_a$SN$_b$ structure (see Fig.~\ref{fig:schema2}b), three
regions are connected to each other: (i) normal electrode N$_a$ at
coordinate $x<0$; (ii) superconducting region S for $0<x<R$, and
(iii) normal electrode N$_b$ for $x>R$.

The two-component wave-function in electrode N$_a$ takes
the form
\begin{eqnarray}
\nonumber \psi_a(x)&=&\left(\begin{array}{c}
1\\0\end{array}\right) \exp{\left(ik_F x\right)} +s_{aa}^{eh}
\left(\begin{array}{c} 0\\1\end{array}\right)
\exp{\left(ik_F x\right)}\\
&+&s_{aa}^{ee}\left(\begin{array}{c} 1\\0\end{array}\right)
\exp{\left(-ik_F x\right)} .
\end{eqnarray}
The two-component wave-function in electrode S takes the form
\begin{eqnarray}
&&\psi_S(x)=c\left(\begin{array}{c} u_0\\v_0\end{array}\right)
\exp{\left(ik_F x\right)} \exp{\left(-x/\xi\right)}\\
\nonumber
&+& d\left(\begin{array}{c} v_0\\u_0\end{array}\right)
\exp{\left(-ik_F x\right)} \exp{\left(-x/\xi\right)}\\
\nonumber
&+& c'\left(\begin{array}{c} u_0\\v_0\end{array}\right)
\exp{\left(-ik_F (x-R)\right)} \exp{\left((x-R)/\xi\right)}\\
\nonumber
&+& d'\left(\begin{array}{c} v_0\\u_0\end{array}\right)
\exp{\left(ik_F (x-R)\right)} \exp{\left((x-R)/\xi\right)}
,
\end{eqnarray}
and in electrode N$_b$ it is given by
\begin{eqnarray}
\nonumber \psi_b(x)&=& s_{ab}^{eh} \left(\begin{array}{c}
0\\1\end{array}\right)
\exp{\left(-ik_F (x-R)\right)}\\
&+&s_{ab}^{ee} \left(\begin{array}{c} 1\\0\end{array}\right)
\exp{\left(ik_F (x-R)\right)} \nonumber .
\end{eqnarray}

The amplitudes are specified by matching the electron and hole
wave functions and their derivatives at both boundaries. They
describe Andreev reflection ($s^{eh}_{aa}$), normal reflection
($s^{ee}_{aa}$), crossed Andreev reflection ($s^{eh}_{ab}$) and
elastic cotunneling ($s^{ee}_{ab}$), as well as transmission in
the superconductor with/without branch crossing of evanescent
states localized at the left/right interfaces (coefficients $c$,
$d$, $c'$ and $d'$). Under standard assumptions, the zero
frequency noise cross-correlations (the time integral of the
real time crossed correlator) are given by\cite{ref-on-noise,Blanter}
$S_{a,b}=\sum_{k,l} S_{a,b}^{(k,l)}$, where $k,l$ label
electrodes $N_a$ and $N_b$, and where
\begin{widetext}
\begin{eqnarray}
\label{noise_cross}
S_{a,b}^{(k,l)}&=&
\int_{0}^{2\pi}
\frac{d(k_{F}R)}{2\pi} 
\sum_{\alpha,\beta,\gamma,\delta \in \{ e,h\}} 
\frac{q_{\alpha}q_{\beta}}{h}
\int d\omega
A_{k\gamma,l\delta}(a\alpha)A_{l\delta,k\gamma}(b\beta)
f_{k\gamma}(1-f_{l\delta})
\end{eqnarray}
\end{widetext}
where Greek indices denotes the nature ($e$ for electrons, $h$ for
holes) of the incoming/outgoing particles with their associated
charges $q_\alpha$, while Latin symbols $l$, $k$ identify the
leads. $f_{k\gamma}$ is a Fermi function for particles of type
$\gamma$ in reservoir $k$ at energy $\hbar\omega$. The matrix
\begin{equation}
A_{k\gamma,l\delta}(i\alpha)\equiv
\delta_{ik}\delta_{il}\delta_{\alpha\gamma}\delta_{\alpha\delta}-
  {{s}^{\alpha\gamma}_{ik}}^{\dagger} s^{\alpha\delta}_{il},
\end{equation}
enters the definition of the current operator, and contains all the information
about the scattering process which were described above.

Note that the noise cross-correlations in Eq.
(\ref{noise_cross}) are averaged over the Fermi phase factor
$k_{F}R$ accumulated  while traversing the
superconductor, which is a standard procedure for evaluating non
local transport through a superconductor
\cite{Melin-Feinberg-PRB}.

\begin{figure}
\includegraphics [width=.7 \linewidth]{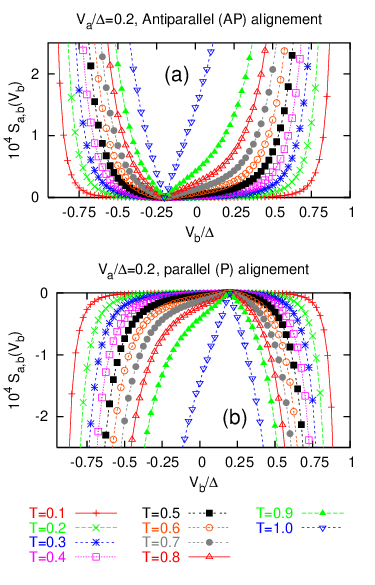}
\caption{Noise $S_{a,b}(e V_b/\Delta)$ of a F$_a$SF$_b$
three-terminal structure with ferromagnetic electrodes F$_a$ and F$_b$
with antiparallel spin orientations (a), and with parallel spin
orientations (b), for interface transparencies ranging from
$T=0.1$ to $T=1$. The voltage on electrode F$_a$ is fixed to
$V_a/\Delta=0.2$. The distance $R$ between the contacts is such
that $\exp{(-R/\xi_0)}=10^{-2}$, with $\xi_0$ the BCS coherence
length at zero energy. The dependence of the coherence length on
energy is supposed to have the BCS form for a ballistic
superconductor. The noise spectra on this figure are obtained from
microscopic Green's functions. \label{fig:Fig34} }
\end{figure}

\subsection{Comparison between the two methods}

 We carried out systematically all calculations 
with the two approaches (microscopic Green's functions and BTK) in order
to determine which are the generic feature of the noise
cross-correlation spectra. We find such model-independent variations of the
noise cross-correlation spectra in the limit of small interface
transparencies\cite{Bignon} and for highly transparent interfaces.
The two approaches do not coincide at the cross-over between small and
large interface transparencies because different
asumptions are made about the geometry.
We complement these numerical results by
an explanation from the Anatram-Datta\cite{ref-on-noise} noise
cross-correlation formula of why noise cross-correlations are positive for
highly transparent interfaces and at small bias.

For microscopic Green's functions we calculate the non
local conductance and the noise cross-correlations of a
single channel contacts, in the spirit of the approach by
Cuevas {\it et al.} \cite{Cuevas}
for a S-S break junction, and averaged over the microscopic
Fermi phase factors.  The electrons (or pairs of electrons)
tunneling through such weak links have to reduce in size down to atomic
dimensions in order to traverse the junction, resulting as in optics
in diffraction. In this approach,
the tunneling Hamiltonian transfers electrons from one point in the 
normal metal lead to one point in the superconductor. The contribution
of wave functions from all momenta in the normal metal are extracted to 
tunnel
into a point in the superconductor therefore transferring to all momenta 
states in the latter (and vice versa). On the opposite, in the 1D BTK model, 
a single (transverse) mode on the normal side is converted into a single mode in the 
superconductor. Diffraction effects are absent for
the one dimensional BTK model. As an example, for perfect interfaces, a
hole cannot be forward-scattered in the 1D BTK model. 
A natural interpretation of this effect for the
1D BTK model\cite{Duhot-Melin} is
momentum conservation for perfect transmission. 

\section{Results}
\label{sec:results}

\subsection{Half-metal/superconductor/half-metal structures}

The dominant transmission channel can be controlled by the
relative spin orientation in the case of half-metals (containing
only majority spin electrons). More precisely, as mentioned in the
Introduction, two processes contribute to the non local
conductance in this case
(see Fig.~\ref{fig:schema1}): elastic cotunneling
(transmission of a spin-up electron from electrode F$_b$ as a
spin-up electron in electrode F$_a$) and crossed Andreev
reflection (transmission of a spin-up electron from electrode
F$_b$ as a hole in the spin-down band in electrode F$_a$ with a
pair left in the superconductor). The dominant non local transport
channel can be selected by the relative spin orientation of
half-metals: elastic cotunneling in the parallel (P)
alignment and crossed Andreev reflection in the antiparallel (AP)
alignment.

The resulting noise cross-correlations are shown on
Fig.~\ref{fig:Fig34} for half-metals in the parallel
(Fig.~\ref{fig:Fig34}a) and antiparallel (Fig.~\ref{fig:Fig34}b)
alignments and for different values of the transmission coefficients
ranging from tunnel to highly transparent interfaces. We obtain a
characteristic sign of noise cross-correlations for the two spin
orientations (negative cross-correlations for elastic
cotunneling, and positive cross-correlations for crossed Andreev
reflection), and a characteristic dependence on the voltage $V_b$
at fixed $V_a$.

\begin{figure}[h]
\includegraphics[width=.8 \linewidth]{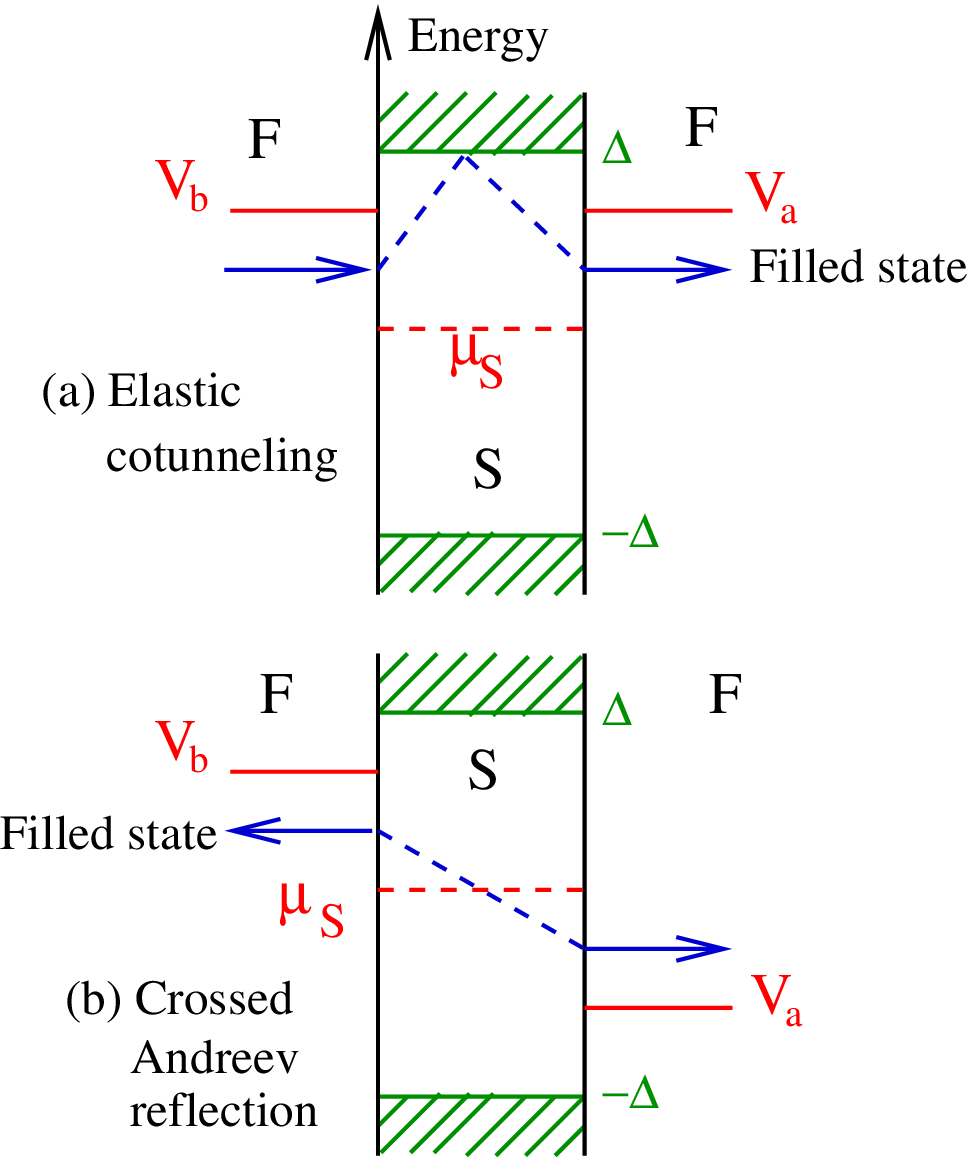}
\caption{Pauli blocking as exemplified in (a)
electron-cotunneling setup and (b) for a crossed Andreev
reflection setup. For (a), an electron incoming from F$_b$
cannot be transmitted in F$_a$ due to Pauli blocking for
$V_b=V_a$. For (b) a pair from the superconductor cannot split
in electrodes F$_a$ and F$_b$ if $V_b=-V_a$.
} \label{Fig:pauli}
\end{figure}

For each setup (antiparallel/parallel alignment), there is a
special value of voltage where the noise cross-correlations vanish.
For instance, the vanishing
of the noise cross-correlations occurs at $V_{b}=-V_{a}$
for the antiparallel configuration. For
half-metals, a (singlet) Cooper pair cannot be transmitted as a
whole in the right or left lead because of the opposite spin
orientation of the two leads. The two electrons of a Cooper pairs have then
to end up into two electrodes with different chemical
potentials, and one has then determine whether the incoming  and
outgoing states are available (see Fig.~\ref{Fig:pauli}).
For the value of voltages
$V_{b}=-V_{a}$, electrons with opposite
energies with respect to the superconducting chemical potential
are not available, and the same is true for pairs of holes with
opposite energies. Therefore the cross-correlations vanish for
this value of voltage, but they increase gradually for voltages
$V_{b}>-V_{a}$ or $V_{b}<-V_{a}$ as Pauli blocking effects
disappear.

\begin{figure}
\includegraphics [width=.8 \linewidth]{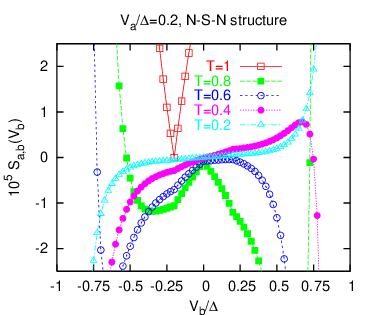}
\caption{Noise $S_{a,b}(e V_b/\Delta)$ of a N$_a$SN$_b$
three-terminal structure with normal electrodes N$_a$ and N$_b$ for
moderate and high interface transparencies (normal transmissions
$T=0.2,0.4,0.6,0.8,1$), within microscopic Green's functions. The
tunnel limit\cite{Bignon} with $S_{a,b}(eV_b/\Delta)$ linear in
$V_b$ in the window $-V_a<V_b<V_a$ is recovered for small $T$.
The distance $R$ between the contacts is such
that $\exp{(-R/\xi_0)}=10^{-2}$, with $\xi_0$ the BCS coherence
length at zero energy.
\label{fig:NSN-green}}
\end{figure}

\begin{figure}
\includegraphics [width=.8 \linewidth]{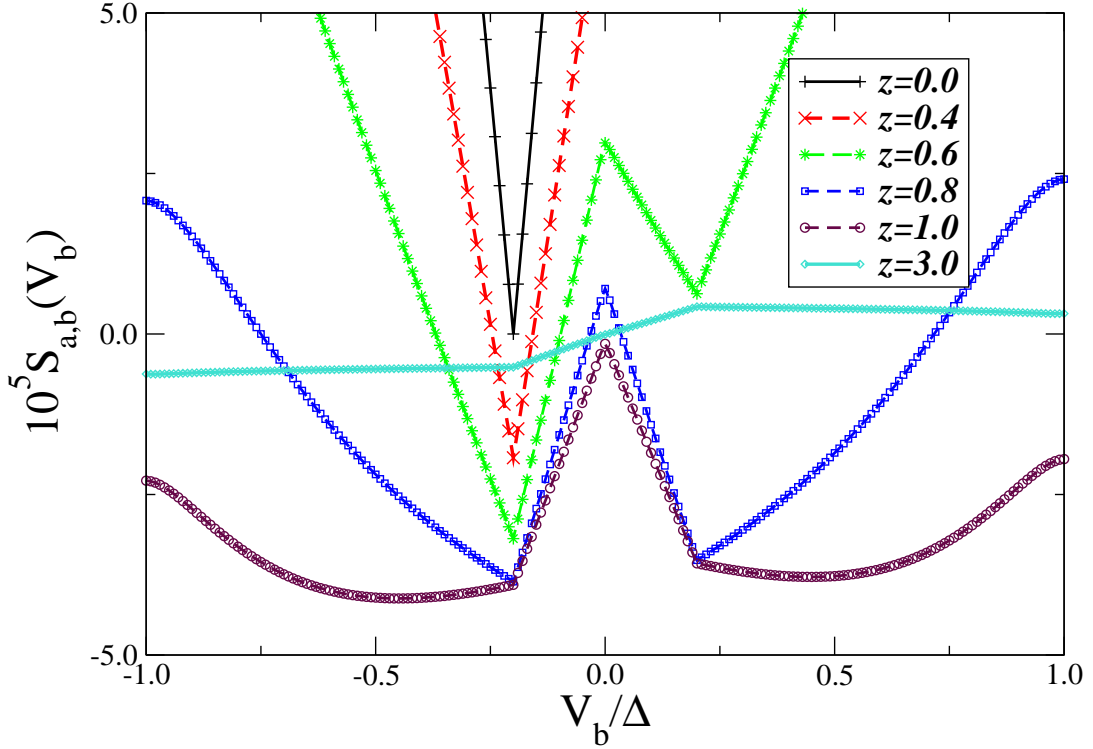}
\caption{Noise $S_{a,b}(e V_b/\Delta)$ of a N$_a$SN$_b$
three-terminal structure with normal electrodes N$_a$ and N$_b$ from
transparent ($z=0.0$) to tunneling ($z=3.0$) interfaces,
from the BTK approach. The tunnel behavior\cite{Bignon} with
$S_{a,b}(eV_b/\Delta)$ linear in $V_b$ in the window
$-V_a<V_b<V_a$ is recovered for large $z$ in the tunnel limit.
The distance $R$ between the contacts is such
that $\exp{(-R/\xi_0)}=10^{-2}$, with $\xi_0$ the BCS coherence
length at zero energy.
\label{fig:NSN-btk}}
\end{figure}


\subsection{Normal metal/superconductor/normal metal structures}

We consider now a one dimensional N$_a$SN$_b$ structure
within the Green's function approach as well as within the one dimensional BTK 
approach.  The dependence of the noise cross-correlations $S_{a,b}(V_a,V_b)$
on $eV_b/\Delta$ 
for a fixed $eV_a/\Delta$  at zero temperature is shown in
Figs.~\ref{fig:NSN-green} and \ref{fig:NSN-btk},
for tunnel interfaces to highly transparent interfaces.
For the BTK approach, the scattering coefficients $s^{ee}_{aa}$, $s^{eh}_{aa}$, $s^{eh}_{ab}$,
and $s^{ee}_{ab}$ are parameterized by a single interface
parameter $z$ (for this symmetric system), $z\equiv
2mH/\hbar^{2}k_{F}$, where $m$ is the band mass, and $H$ is the strength
of the delta potential at the interface. The parameter
$z$ is related to the transmission probability in the normal state
\begin{equation}
z^{2}=\frac{1-T_{0}}{T_{0}}
\end{equation} 
The expected Bignon {\it et al.}\cite{Bignon} tunnel limit for noise
cross-correlations is recovered in both cases. In this case,
the noise cross-correlations vary almost
linearly if the voltage $V_b$ is smaller than $V_a$ (in absolute
value) are recovered for small interface transparencies, in both plots.
In the other limiting case of high transparencies,
the overall shape of the curves is of an
inverted pyramid with its apex at $V_{b}=-V_{a}$. As the transparency
is increased the base of the pyramid enlarges, and the height
decreases.
Similar predictions are obtained in the two approaches
both in the limits of 
small and large interface transparencies.

\subsection{Interpretation}
\begin{figure}[t]
\includegraphics [width=.9 \linewidth]{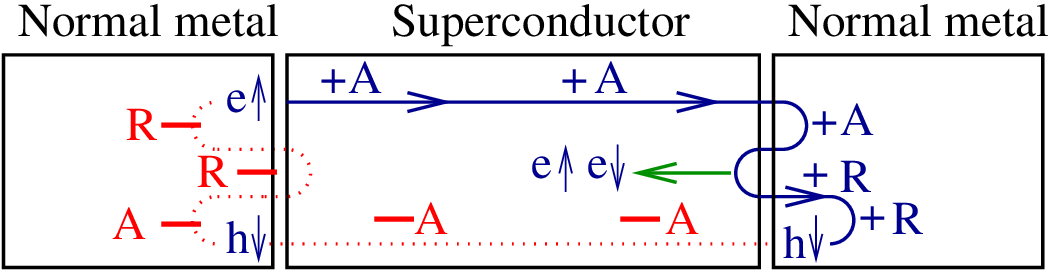}
\caption{Schematic representation of the process
contributing to the noise cross-correlations of 
a N$_a$SN$_b$ structure. 
This process in noise cross-correlations
corresponds to crossed Andreev reflection: propagation forward in time
of an electron from the
left normal metal converted as a hole in the right normal metal
with a pair transmitted in the superconductor,
and the reverse process backward in time. Propagation forward (backward) in 
time are denoted by ``+'' (``-'') on the figure and by solid blue 
(dashed red) lines. We have shown on the figure the pair transmitted
inside the superconductor for propagation forward in time, but, for
clarify, we have not represented the corresponding process for
propagation backward in time. Similarly we have shown the ``advanced'' (A)
and ``retarded'' (R) labels for forward propagation in time.
 \label{fig:doubleAR_new} }
\end{figure}
For high interface transparencies, it is well established
\cite{Melin-Feinberg-PRB,Duhot-Melin,DM-EPJB} that the crossed
conductance is {\it negative}, with the same sign as normal
electron transmission from one electrode to the other across the
superconductor. 

Let us now consider noise crossed correlations. 
We show that positive noise correlations are expected from
the Anatran-Datta \cite{ref-on-noise} noise formula.
We focus 
 on the case of a small applied voltage $eV_{b}\ll\Delta$.
First if $k$ and $l$ in Eq.~(\ref{noise_cross})
both belong to the same electrode N$_a$ we find
\begin{eqnarray}
&&A_{a\gamma,a\delta}(a\alpha)A_{a\delta,a\gamma}(b\beta)
=\\
&&\left(\delta_{\alpha,\gamma}\delta_{\alpha,\delta}
-\overline{s}_{a\gamma;a\alpha}s_{a\alpha;a\delta}\right)
\left(-\overline{s}_{a\delta;b\beta}s_{b\beta;a\gamma}\right)
.
\end{eqnarray}
At zero temperature, the terms containing
$\delta_{\alpha,\gamma}\delta_{\alpha,\delta}$ do not contribute to 
noise cross-correlations because of the Fermi occupation factors
in Eq.(\ref{noise_cross}).
For high transparencies, the term
$\overline{s}_{a\gamma;a\alpha}s_{a\alpha;a\delta}$
encodes local Andreev reflection changing electrons into holes, with therefore
$\alpha=-\gamma=-\delta$. Taking into account the four $s$-matrix 
coefficients and the prefactor containing the sign of the transmitted carriers,
we deduce that this contribution to noise cross-correlation
is vanishingly small.
It can be shown by similar arguments that the term in $S_{a,b}^{(k,l)}$
with $k$ and $l$ labeling the different electrodes N$_a$
and N$_b$ respectively leads to positive noise cross-correlations
for high interface transparencies,
in agreement with Fig.~\ref{fig:NSN-btk}. More precisely,
in terms of scattering matrices, advanced
and retarded propagations correspond to the elements of the
scattering matric $s$ or to their complex conjugate respectively.
To the advanced and retarded labels are added here the labels
encoding propagation forward or backward in time for the two branches
of the Keldysh contour (see Sec.~\ref{sec:MGF}).
Let us consider the term in $S_{a,b}^{(k,l)}$
with $k$ and $l$ labeling the different electrodes N$_a$
and N$_b$ respectively.
We deduce that, for this term, propagation forward
in time 
is the product of an advanced non local propagation 
from N$_b$ to N$_a$,
and a retarded local Andreev process at the interface SN$_a$.
Combining the advanced and retarded terms,
propagation forward in time has the net result of
changing a spin-up electron
from electrode N$_a$ as a hole in the spin-down band in electrode $N_b$,
with a pair left in the superconductor
(as shown on Fig.~\ref{fig:doubleAR_new}), which corresponds to
crossed Andreev reflection with positive noise crossed correlations
(see Fig.~\ref{fig:NSN-btk}).
Pauli blocking is the same as
for crossed Andreev reflection in a F$_a$SF$_b$ structure in the antiparallel
alignment. This picture holds for highly transparent interfaces at low energy.
\section{Consequences for interactions in the leads
(phenomenological approach)}
\label{sec:interactions}

Before concluding, we discuss briefly the inclusion of Coulomb interactions
within the (one dimensional) leads connected to the superconductor.
In normal metal junctions, it is known\cite{ingold} that
electron-electron interactions change qualitatively the current voltage 
characteristics. Such Luttinger liquid behavior 
can be approached from the strong interaction limit, 
or alternatively from the weak interaction limit\cite{fisher_glazman,glazman}. In the latter,
a perturbative renormalization group (RG) scheme
was used to derive the renormalized scattering 
matrix coefficients. 
The interaction-induced dependence of the transmission
probability on energy $\omega$ reads\cite{glazman}:
\begin{equation}
\label{eq:TE}
T(\omega)=\frac{T_{0}|\frac{\omega}{D_{0}}|^{\alpha}}
{1-T_{0}(1-|\frac{\omega}{D_{0}}|^{\alpha})}
\end{equation}
where $T_0$ is the transparency in the absence of interaction,
$0<\alpha<1$ quantifies the electron-electron interaction
strength ($\alpha=0$ corresponds to no interactions),
and $D_0$ is a high-energy cut-off
determined by the energy bandwidth of the electronic states.

For normal metal/superconductor interfaces, this approach was adapted\cite{titov,man} 
to predict that electron-electron interactions induce a suppression of the Andreev
conductance with a non-ideal transparency. However, the Andreev conductance
remains unaffected by interactions for a purely
transparent NS interface. 

\begin{figure}[h]
\includegraphics [width=.7 \linewidth]{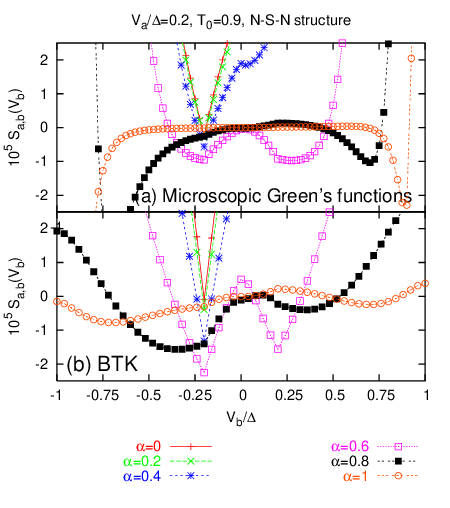}
\caption{Noise $S_{a,b}(e V_b/\Delta)$ of a N$_a$SN$_b$
three-terminal structure with normal electrodes N$_a$ and N$_b$, for an
interface transparency $T=1$ and an interaction parameter $\alpha$
ranging from $0$ to $1$. We use $D_0/\Delta=100$ for the parameter
$D_0$ in Eq.~(\ref{eq:T0}). (a) and (b) correspond to microscopic
Green's functions and to the BTK approach respectively.
The distance $R$ between the contacts is such
that $\exp{(-R/\xi_0)}=10^{-2}$, with $\xi_0$ the BCS coherence
length at zero energy.
\label{fig:alpha-Green} }
\end{figure}
\begin{figure}[h]
\vskip 1.0in
\includegraphics [width=.7 \linewidth]{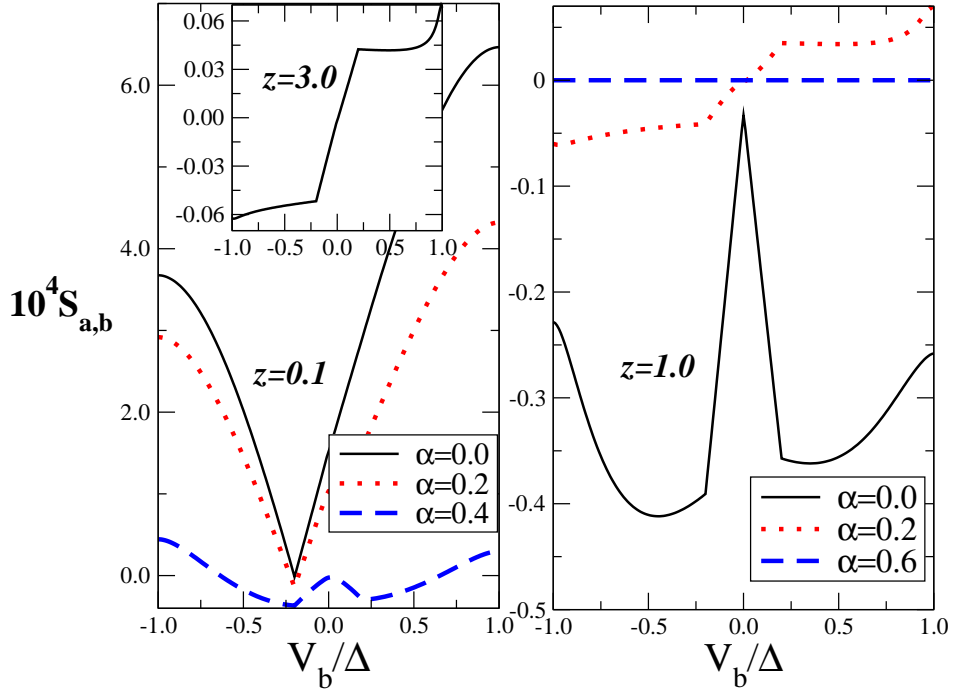}
\caption{The same as Fig.~\ref{fig:alpha-Green} but with
$z=0.1$ (almost transparent) while the right panel is for
$z=1.0$ (semi-transparent). In the inset of the left panel the plot
for the tunneling limit coincides with that of
Ref.[\onlinecite{Bignon}] (linear behavior in between
$-V_{a}<V_{b}<V_{a}$). 
The distance $R$ between the contacts is such
that $\exp{(-R/\xi_0)}=10^{-2}$, with $\xi_0$ the BCS coherence
length at zero energy.
\label{fig:alpha-BTK}}
\end{figure}

For the microscopic Green's function approach of a three terminal 
N$_a$SN$_b$ device, 
the bare transmission amplitude is thus replaced by its energy dependent value 
in order to obtain the noise cross correlations in the presence of interactions
in the leads. 

For the BTK approach,
we follow the phenomenological treatment of Ref. \onlinecite{man}, 
which amounts to replacing $T_0$ by $T(\omega)$ to account for the presence
of interactions. The interaction parameter is changed to $z\to z_{ee}$:
\begin{equation}
z_{ee}^{2}=|\frac{\omega}{D_{0}}|^{-\alpha}
\frac{1-T_{0}}{T_{0}}=|\frac{\omega}{D_{0}}|^{-\alpha}z^{2}
\label{eq:T0}
\end{equation}
When $z=0$ (i.e., $T_{0}=1$), the result $z_{e-e}=0$ implies that for a transparent interface
electron-electron interactions have no effect on electronic
transport\cite{titov,man}.
In the BTK approach $z_{ee}$ is simply 
inserted in the expression for the scattering matrix coefficients, and the noise
cross correlations are computed subsequently. 
We discuss briefly in the conclusion 
the limitations of these approaches (for the Green's function and the BTK situation).  

In Fig.~\ref{fig:alpha-Green}a (Fig.~\ref{fig:alpha-Green}b)
the parameter $\alpha$ is gradually switched on 
to the full interaction value ($\alpha=1$) for  
the Green's function calculation (for the BTK approach). 

For almost transparent contacts
(Fig.~\ref{fig:alpha-BTK}, left panel) we obtain a change from
completely positive crossed correlations throughout the range to
alternating between negative and positive as $\alpha$ is
increased. The weak interactions results are analogous to 
their non-interacting counterparts: single valley at $V_{b}=-V_{a}$. 
However for strong interactions  
two valleys are located at $V_{b}=\pm V_{a}$.
Further increase 
of interactions
leads to a gradual smoothing of these features, 
eventually leading to the result for 
tunnel interfaces \cite{Bignon}.

This is in contrast with the case of semi-transparent contacts
wherein noise cross-correlations which are uniformly negative
(as in Fig.~\ref{fig:alpha-BTK}, right panel) turn positive as
interactions are increased in the positive voltage range. 
Indeed, for this situation, interactions change a transparent 
contact into a tunneling
contact. 

\section{Conclusions}
\label{sec:conclu}

Experiments on non local Andreev reflection \cite{Cadden,Russo,Beckmann}
have already been achieved, and they have generated a lot of excitement in the
mesoscopic 
physics community. 
Noise cross-correlation experiments in normal metal/superconducting hybrid
structures
are the next on the list and experiments are under way. Such experiments could
allow to 
observe the splitting of Cooper pairs from a superconductor into two distinct
normal metal leads. 

This would constitute a milestone toward  probing entanglement in 
the context of nanophysics. A renewed theoretical interest into such questions, 
taking into account experimental constraints (the separation between contacts
is large compared to the superconducting coherence length) is simply warranted. 

In the beginning of this work, we computed 
the noise cross-correlation
spectra as a function of voltage with varying interface
transparencies for half-metal/superconductor/half-metal systems.
For oppositely polarized half-metals noise cross-correlations are
positive because crossed Andreev reflection is then the
only possible channel of non local transport. Crossed Andreev reflection leads to positive
cross-correlations as electron and holes detected at separate
electrodes arise from the same Cooper pair in the superconductor.
However, for half-metals polarized in the parallel alignment,
cross-correlations are negative because elastic
cotunneling is the only means of transport. 
Elastic cotunneling leads to negative
cross-correlation since electrons detected at the separate
electrodes do not share any information.
These results, which display the physics of Pauli Blocking, have 
an obvious interpretation in the tunneling limit.
When the interface transmission is increased, we find that the 
the overall sign of the noise correlations is unchanged. However, 
both for the parallel and anti-parallel configuration, the amplitude of 
noise correlations are much more sensitive to the applied voltage, 
and the slope (in absolute value) around the vanishing of such correlations 
is increased.   

But the most important part of this work concerns the results for 
normal metal leads, which are more accessible experimentally, 
but for which the splitting of Cooper pair cannot be achieved by 
projecting the spin. 
As a first check, we recover the tunneling limit results\cite{Bignon}, 
where the sign of cross correlations goes from positive 
to negative by varying the voltages attributed to each lead, and thus 
favoring cross Andreev processes or electron cotunneling processes.
As the transparency is gradually increased, we observe a crossover to positive 
correlations. This result 
is especially robust close to ideal transparency, which
constitutes the main result of this paper. This is especially surprising 
in the light of previous work on the non-local conductance.

We conclude that a negative crossed conductance is compatible with positive noise
cross-correlations for highly transparent interfaces
in N$_a$SN$_b$ structures and this is a 
consequence of the different underlying process which contribute to the noise
and in the non local conductance. For the  noise cross-correlations,
an electron from N$_b$ is converted into a hole in N$_a$ with a pair transmitted
in the superconductor, as in non local (cross) Andreev reflection.
For the non local conductance, a pair of electrons from N$_b$
tunnels through the superconductor into electrode N$_a$ effectively in the
form of two-electron cotunneling. 
This means that for transparent interfaces, the negative crossed conductance
is not expected to be a signature of entanglement whereas 
the positive noise crossed correlation signal provides an evidence 
for the presence of entangled electrons between the two leads.

We found it informative to address the issue of 
interactions, if the normal metal leads are one dimensional. 
Rather than providing a detailed derivation of the renormalization procedure
for this system with two interfaces, we chose a 
phenomenological approach\cite{man} which ``only'' renormalizes
the hopping amplitudes/interaction parameters at the two interfaces, 
rather than treating the two leads system as a whole.  
We find that for near ideal interfaces, interactions reduce the amplitude of 
the positive noise cross correlation signal and can even reverse its sign for 
strong interactions. For intermediate to low interfaces, upon renormalization
the result coincides with the tunneling limit as it should be. 
Granted, these constitute preliminary results, because in particular the 
renormalization of the transmission phase is not described
rigorously by renormalization group equations, as in
Refs.[\onlinecite{titov,takane}]. However, as far as the current for a single 
interface is concerned,
results from the exact works are reproduced with this
phenomenological approach, as seen from Refs.
[\onlinecite{man,takane}]. One could expect that the separation between 
the two normal metal/superconductor interfaces could complicate things 
(compared to a single interface), 
but many round trips between the two interfaces are
prohibited by our assumption of a large contact separation 
(compared to the coherence length).   
Moreover, for single interfaces, the renormalization of the Andreev
reflection {\it phase} is predicted\cite{titov} to affect 
only transport through geometries containing 2 superconductors (Josephson effect).
Nevertheless, the renormalization treatment of the full two interface
problem would constitute an extension of the present work.

Effects associated with one dimensional leads could be experimentally probed 
when connecting a superconductor to carbon nanotube wires, which
could exhibit Luttinger liquid behavior. Experiments with a single nanotube
split into two leads by an overlapping superconducting contact have been 
recently achieved in Ref. \onlinecite{cleusiou}.

A connection between the work of 
Ref. \onlinecite{levy-yeyati} and the present situation with interacting 
one dimensional leads can be envisioned, due to the fact that 
the physics of dynamical Coulomb blockade\cite{ingold} is intimately 
tied to the physics of tunneling between Luttinger liquids
(excitation of ``many'' bosonic modes during an electron tunneling event).  
Further work dealing with the finite frequency spectrum of noise
cross-correlations, possibly in the presence of such
interactions\cite{levy-yeyati} would constitute a relevant
extension of this work. 
 


\section{Acknowledgments}
R.M. acknowledges a fruitful discussion with D. Feinberg on
the BTK approach to non local transport.
T.M.  acknowledges support from ANR grant ``Molspintronics'' from
the French ministry of research.

\end{document}